\documentstyle[aps,prl,multicol,psfig,graphics]{revtex}
\tighten
\begin{document}

\title{Dynamical charge susceptibility in layered cuprates: \\
       the influence of screened inter-site Coulomb repulsion}

\author{M. Eremin$^1$, I. Eremin$^{1,2}$, and S. Varlamov$^{1,3}$}
\address{$^1$Physics Department, Kazan State University
Kremlyovskaya 18, 42008 Kazan, Russia}
\address{$^2$ Freie Universit\"at Berlin, Arnimallee 14, D-14195 Berlin,
Germany}
\address{$^3$ Cottbus Technical University, D-03013 Cottbus, Germany}

\date{\today}
\maketitle

\begin{abstract} 
The analytical expression for dynamical charge susceptibility 
in layered cuprates has been derived in the frame of singlet-correlated 
band model beyond random-phase-approximation (RPA) scheme.
Our calculations performed near optimal doping regime show that there is
a peak in real part of the charge susceptibility $\chi({\bf q},\omega)$ 
at {\bf Q} = ($\pi$, $\pi$) at strong enough inter-site Coulomb repulsion.
Together with the strong maximum in the Im $\chi({\bf Q},\omega)$ 
at 15 meV it confirms the formation of low-energetic plasmons or 
charge fluctuations. This provides a jsutification that these excitations
are important and together with a spin flcutuations can 
contribute to the Cooper pairing in layered cuprates.
Analysing the charge susceptibilitiy with respect to an instability 
we obtain a new plasmon branch, $\omega_{\bf q}$, 
along the Brillouin Zone. In particular, we have found that 
it goes to zero near {\bf Q}$_{CDW} \approx (2\pi/3, 2\pi/3)$.
\end{abstract}


\vspace*{0.2cm}

\begin{multicols}{2}
\section{Introduction}

The phenomenon of high-temperature superconductivity can not be completely
understood without clarification of 'pseudogap' features
seen by many experimental techniques in the normal state of underdoped 
cuprates. There are few informative
reviews and papers devoted to this problem( see for example
 \cite{timusk,loram,marki,kivelson}). One of the most reasonable 
point of view is that a pseudogap in the density of states shows up due to an
instability in two-dimensional copper-oxygen planes.
The important question to answer is what is the 
physical origin of this possible instability in layered cuprates.
In a series of works Di Castro and co-workers(see for example \cite{grilli}) 
in a frame
of the phenomenological quantum critical point (QCP) hypotethis have 
suggested to explain the pseudogap as a result of the formation of the 
incommensurate charge density wave (ICDW).
In recent paper the pseudogap phenomena has been
considered in a context of a new hidden order 
parameter with orbital antiferromagnetism
(special case of d-wave wave order parameter) \cite{chakrav} .
Constructing the integral equation for the extended charge density waves 
instability in a frame t-J Hamiltonian proposed by 
Anderson \cite{anderson} it is easy to prove 
that for $q=(\pi,\pi)$ the order parameter for charge instability
indeed has a d-wave symmetry. However, the order parameter is 
imaginary\cite{ieremin} and, 
therefore, is not a usual charge density waves. This reminds the 
old problem of the itinerant currents studied many years ago in 
semiconductors \cite{kopaev,rice}.

Perhaps, the most convincing evidence about the instability 
in layered cuprates can be obtained directly within the analysis of the
dynamical charge and spin susceptibilities. In contrast to the 
spin susceptibility the role played by the instabilities in the 
charge channel with respect to the pairing and pseudogap formation 
are not clear. Moreover, the form of the charge susceptibility beyond
RPA scheme was much less investigated. In particular, there are no agreement
about its analytical form if the 
effect of strong electron correlations is taken into account 
(for example, compare the results in Refs. \cite{plakida,zeyher,santoro}.
Therefore, the theoretical investigation in this direction becomes 
very actual.

Here, using the singlet-correlated
band model (\cite{prberemin,eve} and references therein) which under
simplified assumption about energy dispersion and screened Coulomb repulsion 
is equivalent to the $t-J$ model \cite{zhang} using the 
projecting Hubbard-like operators \cite{hubbard} 
we derive the expression for the dynamical charge 
susceptibility beyond RPA-scheme. 
We analyze the obtained expression numerically and find that 
the charge susceptibility shows a peak at {\bf Q} =($\pi$, $\pi$) similarly
to a spin susceptibility only if one takes into account the strong enough
inter-site Coulomb repulsion. 
Investigating the denominator of the charge
susceptibility we have found the possible dispersion curve of 
CDW-like excitations $\omega_{q}$, for optimally doped cuprates.
 
\section{Model Hamiltonian and Background}

The Hamiltonian in the singlet-correlated band model in terms of Hubbard-type 
projecting
operators, $\Psi_{i}^{\alpha,\beta}=|i,\alpha><i,\beta|$ reads as
\begin{equation}\label{hubbard}
H=\sum t_{ij} \Psi_{i}^{pd,\sigma} \Psi_{j}^{\sigma,dp}+
\sum J_{ij}[(S_iS_j)-\frac{n_in_j}4]+
\sum G_{ij}\delta _i\delta_{j} \\  
\end{equation}
where $t_{ij}$ is a hopping integral and $J_{ij}$ is the superexchange 
constant of the  copper spins, $\sigma=\pm 1/2$. 
Symbol $pd$-corresponds to a  Zhang-Rice 
singlet where one  hole  is placed on copper and second distributed on 
neighboring oxygen sites \cite {zhang} and carrying spin 
states, respectively. $g_{ij}$ is a screened inter-site Coulomb 
repulsion of the doped
holes. $1+\delta_{i}= \sum \Psi_{i}^{\sigma\sigma} + 2 \Psi_{i}^{pd,dp}$ is
equation that determines the chemical potential.
The $\Psi$ operators obey the specific commutation relation and 
on-site multiplication rules of Hubbard-like operators which can be found 
somewhere \cite{hayn,esv}. The spin and density operators are expressed by 
projecting operators as follows
\begin{eqnarray}\label{density}
S_{i}^{+}&=&\Psi_{i}^{\uparrow, \downarrow}, \,\,\,
S_{i}^{-}=\Psi_{i}^{\downarrow, \uparrow}, \,\,\,
S_{i}^{z}=\frac{1}{2}( \Psi_{i}^{\uparrow,\uparrow}
-\Psi_{i}^{\downarrow,\downarrow}), \nonumber \\
\delta_{i}&=& \Psi_{i}^{pd,dp}- \Psi_{i}^{0,0}.
\end{eqnarray}
Let us now make some preliminary discussion about the nature of density wave 
formation in a frame of our model. We are going to study the objects which
can be described by the following operators
\begin{eqnarray}\label{cdw}
\eta_{q} = \frac{1}{2} \sum A({\bf k,q},\omega)\lbrack \Psi_{k+q}^{pd, \uparrow}
\Psi_{k}^{\uparrow, pd}+\Psi_{k+q}^{pd, \downarrow}
\Psi_{k}^{\downarrow, pd} \rbrack ,
\end{eqnarray}
\begin{eqnarray}\label{sdw}
\xi_{q} = \frac{1}{2} \sum B({\bf k,q},\omega)\lbrack 
\Psi_{k}^{\uparrow, pd}\Psi_{k+q}^{pd, \uparrow}-
\Psi_{k}^{\downarrow, pd}\Psi_{k+q}^{pd, \downarrow} \rbrack ,
\end{eqnarray}
where $A({\bf k, q}, \omega)$ and $B({\bf k, q}, \omega)$ are 
charge- and spin-excitation amplitudes, respectively. 
The frequencies of the harmonic 
motion of these objects can be found in a usual manner
\begin{eqnarray}
i\frac{\partial \eta_{q}}{\partial t} = \lbrack \eta_{q},H \rbrack=\omega_{q}^c \eta_{q}, \quad
i\frac{\partial \xi_{q}}{\partial t} = \lbrack \xi_{q},H \rbrack=\omega_{q}^s\xi_{q}.
\label{motion}
\end{eqnarray}
Real part of $\eta_{q}$ corresponds to usual charge density waves (CDW)
order parameter whereas its
imaginary part is responsible for the charge current 
(JC) formation \cite{rice,shulz}. The latter also 
looks very similar to
the proposed earlier flux phase \cite{affleck}
and orbital antiferromagnetism \cite{chakrav}. 
Real part of $ \xi_{q}$ represents the well-known 
spin density waves (SDW) and its imaginary quantity 
closely related to the so-called spin current (JS) formation
\cite{rice,shulz}. Symmetry aspects of these interesting problem 
has been discussed very recently by Nayak \cite{nayak} and 
therefore we do not touch this problem here. 
One could immediately see from (3) and (4) that the equations for the 
frequency determination of the real and imaginary parts looks
very similar and seem to be a good starting point to study both 
the collective charge and the spin excitations in layered cuprates. 

Calculating the commutators using the simplest  Hubbard1 decoupling 
scheme\cite{hubbard} (for details see the next section) one can find that
$A({\bf k,q,}\omega)$ and $B({\bf k,q,}\omega)$ can be taken in a usual form:
\begin{eqnarray}
A({\bf k,q,}\omega)&=&\frac{1}{\omega_{q}^{c}-\epsilon_{k+q}+\epsilon_{k}}, \nonumber \\
B({\bf k,q,}\omega)&=&\frac{1}{\omega_{q}^{s}-\epsilon_{k+q}+\epsilon_{k}},
\label{assumption}
\end{eqnarray}
and frequencies are determined by the relations
\begin{equation}
1=\frac{1}{2} \sum \frac{t_{k+q}n_{k+q}-t_{k}n_{k}}{\omega_{q}^{c}-
\epsilon_{k+q}+\epsilon_{k}} - \lgroup \frac{J_{q}}{4}-g_{q}\rgroup \sum
 \frac{n_{k}-n_{k+q}}{\omega_{q}^{c}-
\epsilon_{k+q}+\epsilon_{k}}  \\
\label{ocdw}
\end{equation}
\begin{equation}
1=\sum \frac{t_{k+q}n_{k+q}-t_{k}n_{k}}{\omega_{q}^{s}-
\epsilon_{k+q}+\epsilon_{k}} - \frac{1}{2}J_{q} \sum
 \frac{n_{k+q}-n_{k}}{\omega_{q}^{s}-
\epsilon_{k+q}+\epsilon_{k}}
\label{osdw}
\end{equation}
where $n_{k}=\langle \Psi_{k}^{pd, \sigma} \Psi_{k}^{\sigma, pd} \rangle$ 
is a partion number, $\epsilon_{k}= P_{pd}t_{k}$ is an energy
dispersion of the itinerant carriers, 
$P_{pd} = \frac{1+\delta}{2}$ is Hubbard-type bandwidth narrowing factor. 
The Fourier transform of the integrals $t_k$, $J_q$, and $g_q$
are determined by the usual expressions
\begin{eqnarray}
t_k & = &  2t_1 \lgroup \cos k_x + \cos k_y \rgroup +4t_2 \cos k_x \cos k_y
%+  2t_3 \lgroup \cos 2k_x + \cos 2k_y \rgroup 
+\cdots  \\
J_q &  = &  2J_1 \lgroup \cos q_x + \cos q_y \rgroup +4J_2 \cos q_x \cos q_y 
+ \cdots  \\
G_q & = & 2G_1 \lgroup \cos q_x + \cos q_y \rgroup +4G_2 \cos q_x \cos q_y
+ \cdots
\label{fourier}
\end{eqnarray}
where $t_1$($J_1$, $g_1$), $t_2$($J_2$, $g_2$), and $t_3$ refer to the
first, second, and third neighbors respectively. For
simplicity we put the lattice constant equals unity. The choice of parameters
for the hopping integrals is determined by the right Fermi surface topology
and presence of Van-Hove singularity in the vicinity of the Fermi level 
as observed in the experiment \cite{shen}. This satisfies at $t_1$ = 72 meV,
$t_2$ = 0, and $t_3$ = 12 meV. The screened Coulomb repulsion of the doped 
holes and superexchange integral between nearest copper spins were taken 
as $G_1$ = 70 meV and $J_1$ = 135 meV in agreement with first principles
calculation\cite{coulomb} and inelastic neutron scattering experiments
\cite{bourges} respectively.  
  
Regarding Eqs. (\ref{ocdw}) and (\ref{osdw}) we would like to remark two
important features. Both equations determine the conditions of the 
instability of
the system with respect to mentioned types of charge and spin  instabilities. 
In other words  Eqs. (\ref{ocdw}) and (\ref{osdw}) determine the divergences
of charge and spin susceptibilities in the normal state. 
In contrast to
the usual RPA-type of expression the first terms in 
(\ref{ocdw}) and (\ref{osdw}) proportional to the hopping 
integral result
from the  strong electron correlation effects 
due to no double on-site occupancy 
constraint\cite{esv}. It reflects the existence of very large 
on-site Coulomb repulsion leading to a renormalization and enhancement 
of the susceptibilities (or in other words the presence of strong 
electron correlation). This non-Fermi liquid correction was 
firstly deduced by Hubbard and 
Jain\cite{jain} for the spin susceptibility. The second remark is that 
as one can see from Eqs. (\ref{ocdw}) and (\ref{osdw}) the
screened inter-site Coulomb repulsion plays no role for the determining of spin 
dispersion in contrast to charge one.  
Therefore it becomes clear that the 
equations for charge and spin collective excitation are quite 
different in presence of inter-site Coulomb interaction than 
it was widely implied.  However, our present approach is quite 
preliminary and in order to see the difference between charge and spin 
collective excitations more in details we will calculate 
dynamical charge susceptibility and compare its form with spin counterpart
obtained earlier\cite{zavidonov}. 

\section{Dynamical charge susceptibility. Improved decoupling scheme
for the equation of motions}

We start from the definition of the Fourier transform  of the 
amplitude of the density operator
\begin{eqnarray}\label{furcdw}
e_q = \Psi_{q}^{pd,pd}&=&\frac{1}{N}\sum_{k} e_{kq}     \nonumber\\
&=& \frac{1}{2N}\sum_{k} \lgroup \Psi_{k}^{pd,\uparrow} 
\Psi_{k+q}^{\uparrow, pd} + \Psi_{k+q}^{pd,\downarrow} 
\Psi_{k}^{\downarrow, pd} \rgroup
\end{eqnarray}
where the Fourier transform defined as
\begin{equation}\label{furcdw1}
\Psi_{k}^{pd,\uparrow}  = \frac{1}{\sqrt{N}} \sum_{k} \Psi_{j}^{pd,\uparrow}
e^{i{\bf qR_{j}}}. 
\end{equation}
In the next step we derive the equation of motion for the two retarded
Green's functions
\begin{equation}\label{green1}
\omega_{q}^c \langle \langle e_{q} \vert e_{-q} \rangle \rangle_{\omega} 
= \frac{i}{2\pi} \lbrack \Psi_{q}^{pd, pd}, \Psi_{-q}^{pd, pd} \rbrack 
+ \langle \langle \lbrack \Psi_{q}^{pd, pd}, H \rbrack \vert e_{-q}
\rangle \rangle_{\omega}. \\
\end{equation}
and 
\begin{equation}\label{green2}
\omega_{q}^c \langle \langle e_{kq} \vert e_{-q} \rangle \rangle_{\omega} 
= \frac{i}{2\pi} \lbrack e_{kq}, e_{-q} \rbrack 
+ \langle \langle \lbrack \,e_{kq} H \rbrack \vert e_{-q}
\rangle \rangle_{\omega}. \\
\end{equation}
We calculate the commutators using the on-site representation of the operators.
For example for the kinetic part of the Hamiltonian $H_t$,
it has a form
\begin{eqnarray}\label{commut}
&&\lbrack \Psi_{i}^{pd, \uparrow}\Psi_{j}^{\uparrow, pd},H_t \rbrack =  \nonumber\\
&=&\Psi_{i}^{pd, \uparrow} \sum_{l} t_{jl}\lbrace \Psi_{j}^{\downarrow,
\uparrow} \Psi_{l}^{\downarrow, pd} +\lgroup \Psi_{j}^{\uparrow, \uparrow} +
\Psi_{j}^{pd, pd} \rgroup \Psi_{l}^{\uparrow, pd} \rbrack -        \nonumber\\
&-& \sum_{l} t_{lj} \lbrace \Psi_{l}^{pd, \downarrow}
\Psi_{i}^{\downarrow, \uparrow} + \Psi_{l}^{pd, \uparrow} \lgroup
\Psi_{i}^{\uparrow, \uparrow} + \Psi_{i}^{pd, pd} \rgroup \rbrace
\Psi_{j}^{\uparrow pd}.
\end{eqnarray}
This exact result leads, however, to the appearance of  Green's function of
higher order. Therefore we made the decoupling in the few steps. First, using
the equation for the determining of the chemical potential and definitions of 
the density and spin operators determined in previous section one can find 
\begin{eqnarray}\label{simpl1}
\Psi_{j}^{\uparrow \uparrow} + \Psi_{j}^{pd pd} & = & \frac{1+\delta}{2}
+\frac{e_j}{2} + s^{z}_{j} \\
\Psi_{j}^{\downarrow \downarrow} + \Psi_{j}^{pd pd} & = & 
\frac{1+\delta}{2} + \frac{e_j}{2} - s^{z}_{j}
\end{eqnarray}
where $\delta$ is an average number of the doped holes per one unit cell and 
$e_i$ its modulation. The next step is to differ the cases $i=j$ and $i\ne j$. In 
latter the right hand side of (\ref{commut}) can be approximated within the
usual approximation technique
\begin{eqnarray}\label{hub1}
\sum_{l} &t_{jl}& \lbrace \Psi_{j}^{\uparrow,
\downarrow} \langle \Psi_{i}^{pd ,\uparrow} \Psi_{l}^{\uparrow, pd} \rangle +
\frac{1+\delta}{2} \Psi_{i}^{pd, \uparrow}\Psi_{l}^{pd, \uparrow}  \nonumber\\
&+&\lgroup \frac{e_j}{2} + s_{j}^{z} \rgroup
\langle \Psi_{i}^{pd,\uparrow} \Psi_{l}^{\uparrow, pd} \rangle \rbrace 
-\sum_{l} t_{lj} \lbrace \langle \Psi_{l}^{pd, \downarrow} 
\Psi_{j}^{\downarrow, pd} \rangle \Psi_{i}^{\downarrow,\uparrow}   \nonumber\\
&+&\frac{1+\delta}{2} \Psi_{l}^{pd, \downarrow}
\Psi_{j}^{pd, \downarrow}  + \lgroup \frac{e_i}{2} + s_{i}^{z} \rgroup
\langle \Psi_{l}^{pd, \downarrow} \Psi_{j}^{\downarrow, pd} \rangle \rbrace.
\end{eqnarray}
where $s_{i}^{z}$ is zero in the paramagnetic phase.

As for  the case $i=j$ it can be calculated exactly as
\begin{eqnarray}\label{i=j}
\lbrack \Psi_{i}^{pd, pd}, H_t \rbrack 
&=& \sum_{l} t_{il} \lgroup \Psi_{i}^{pd, \uparrow} \Psi_{l}^{\uparrow, pd}+
\Psi_{i}^{pd, \downarrow} \Psi_{l}^{\downarrow, pd} \rgroup    \nonumber\\
&-& \sum_{l} t_{li} \lgroup \Psi_{l}^{pd, \uparrow} \Psi_{i}^{\uparrow, pd}+
\Psi_{l}^{pd, \downarrow} \Psi_{i}^{\downarrow, pd} \rgroup 
\end{eqnarray}
Using the Fourier transformation (\ref{furcdw1}) we can write the commutator
(\ref{commut})
in the equation of motion as follows
\begin{eqnarray}\label{q}
&&\lbrack e_{kq}, H_t \rbrack = P_d \lgroup t_k - t_{k+q} \rgroup e_{kq}  \nonumber\\
&+&\frac{1}{4N} \sum_{k} \lbrack t_{k+q} \lgroup n_{k+q}^{\uparrow} +
n_{k+q}^{\downarrow} \rgroup - t_k \lgroup n_{k}^{\uparrow} +
n_{k}^{\downarrow} \rgroup \rbrack e_q                           \nonumber\\
&+& \frac{2-P_{pd}}{2N} \sum_{k'} \lgroup t_k' - t_{k'+q} \rgroup e_{k'q}.
\end{eqnarray}
The last term in (\ref{q}) still leads to the high order Green's function but 
hopefully it can be reduced to the old one with the help of (\ref{i=j}),
which yields
\begin{equation}\label{reduce}
\frac{\omega_q}{2} \langle \langle e_q | e_{-q} \rangle \rangle = 
\sum_{k} \lgroup t_k - t_{k+q} \rgroup \langle \langle e_{k'q} 
| e_{-q} \rangle \rangle.
\end{equation}
Turning to the superexchange and Coulomb screening parts of the Hamiltonian,
$H_J$ and $H_c$, we notice that operator $\Psi_{i}^{pd,pd}$ commutes with 
these 
terms of the Hamiltonian and therefore $i=j$ case does not contribute
to the equation of motion. This is in contrast to the spin susceptibility where
the additional non-commutation of spin operator $S^{+}_{q}$ led to the 
fact that ordinary Hubbard1 decoupling 
scheme is not enough and additional consideration is required 
\cite{zavidonov}. Strictly speaking it means that although spin and charge
degrees of freedom are still coupled as in the case of the usual Fermi-liquid,
however, the simple relations of RPA-approximation of the usual Fermi liquid 
approach do not hold.
 
For the case of $i\ne j$ in site representation we have
\begin{eqnarray}\label{hj}
\lbrack \Psi_{i}^{pd \uparrow} &\Psi_{j}^{\uparrow pd}&,H_J \rbrack
= \frac{1}{2} \sum_{l} J_{il} \lbrace \Psi_{i}^{pd \downarrow} 
\Psi_{l}^{\downarrow \uparrow} \Psi_{j}^{\uparrow pd} -  
\Psi_{i}^{pd \uparrow} \Psi_{l}^{\downarrow \downarrow} 
\Psi_{j}^{\uparrow pd} \rbrace  \nonumber \\
&-&\frac{1}{2}\sum_{l} J_{jl} \lbrace  \Psi_{i}^{pd \uparrow} 
\Psi_{j}^{\downarrow pd} \Psi_{l}^{\uparrow \downarrow} - 
\Psi_{i}^{pd \uparrow} \Psi_{j}^{\uparrow pd}
\Psi_{l}^{\downarrow \downarrow} \rbrace.
\end{eqnarray}
whereas for the case  $i=j$
\begin{equation}\label{zj}
\lbrack \Psi_{i}^{pd,pd}, H_J \rbrack = 0
\end{equation}
Restricting ourself an absence of long range spin order
we can wright
\begin{equation}\label{hjj}
\lbrack e_{kq}, H_{J} \rbrack = \frac{1}{4N} J_q \lbrace \lbrack 
n_{k}^{\uparrow} + n_{k}^{\downarrow} \rbrack - \lbrack n_{k+q}^{\uparrow}
+n_{k+q}^{\downarrow} \rbrack \rbrace e_q.
\end{equation}
For the Coulomb screening part of the Hamiltonian the result
reads
\begin{equation}\label{hgg}
\lbrack e_{kq}, H_{c} \rbrack = -\frac{1}{2N} g_q \lbrace \lbrack 
n_{k}^{\uparrow} + n_{k}^{\downarrow} \rbrack - \lbrack n_{k+q}^{\uparrow}
+n_{k+q}^{\downarrow} \rbrack \rbrace e_q.
\end{equation}
Substituting (\ref{q},\ref{hjj},\ref{hgg}) into the equation of motion 
(\ref{greenf}) and taking into account the absence of the external 
magnetic field we get finally
\begin{eqnarray}\label{greenf}
\omega_{q} \langle \langle e_{kq} | &e_{-q}& \rangle \rangle= \frac{1}{2\pi}
\lgroup n_{k+q} - n_k \rgroup + P_{pd} \lgroup t_k - t_{k+q} \langle
\langle e_{kq} | e_{-q} \rangle \rangle  \nonumber \\
&+& \lbrack \frac{1}{4} \lgroup t_{k+q}n_{k+q} - t_{k}n_{k} 
\rgroup + \lgroup G_q -\frac{J_q}{4} 
\rgroup \lgroup n_{k+q} - n_{k} \rgroup   \nonumber \\
&+&\frac{\lgroup 2-P_{pd} \rgroup}{2}{\omega_{q}} \rbrack \langle \langle
e_q | e_{-q} \rangle \rangle
\end{eqnarray}
where $n_k = \frac{1}{2}(n_{k}^{\uparrow}+n_{k}^{\downarrow})$.
Performing a sum over {\bf k} we obtain the expression for the dynamical 
charge susceptibility in a form
\begin{equation}
\chi_{ch}({\bf q},\omega) = \frac{\chi_{0}({\bf q}, \omega)}{1+
\lgroup G_q -\frac{J_q}{4}\rgroup \chi_{0}({\bf q}, \omega) + 
\chi_{1}({\bf q}, \omega) - \lgroup 1-\frac{P_{pd}}{2}\rgroup 
z({\bf q}, \omega)}
\label{cds}
\end{equation}
where $\chi_{l}({\bf q}, \omega)$ is an ordinary Pauli-Lindhard response 
function
\begin{equation}
\chi_{0}({\bf q},\omega) = \frac{1}{N} \sum_k \frac{n_{k}-n_{k+q}}{\omega +
i0^{+} - \epsilon_k + \epsilon_{k+q}}. 
\label{lindhard}
\end{equation}
The most interesting $\chi_1({\bf q}, \omega)$ 
and $z({\bf q}, \omega)$ terms describe the contributions due to 
no-double occupancy constraint and projecting nature of Hubbard operators:
\begin{eqnarray}\label{ch1q}
\chi_1({\bf q},\omega)=\frac{1}{2N} \sum_{k} \frac{t_{k}n_{k} -t_{k+q} n_{k+q}}
{\omega +i0^{+} -\epsilon_k + \epsilon_{k+q}},
\end{eqnarray}
\begin{eqnarray}\label{zq}
z_1({\bf q},\omega)=\frac{1}{N} \sum_{k} \frac{\omega + i0^{+}}
{\omega +i0^{+} -\epsilon_k + \epsilon_{k+q}}.
\end{eqnarray}
The comparison of the obtained equation with our preliminary consideration
resulting in Eq. (\ref{ocdw}) shows that important
 $\frac{\lgroup 2-P_{pd} \rgroup }{2} z ({\bf q}, \omega)$ 
term has been appeared on the right hand side of
(\ref{zq}). To our knowledge this term was not pointed out for 
the charge susceptibility so far. It acts as a frequency dependent 
"molecular field" caused by the transfer hopping term of the Hamiltonian 
(\ref{hubbard}).
 
\section{Numerical results and discussion}
\vspace{-1.5cm}
Let us first analyze the behavior of the obtained expression for the 
dynamical charge susceptibility (\ref{cds}). 
In Fig. 1 we present the calculated Re $\chi_{ch}({\bf q}, 0)$ for
the set of parameters described above. The peak around ($\pi$, $\pi$)
reflects the nesting properties of the Fermi surface enhanced by the 
RPA-type of the denominator. The structure of real part of the charge 
susceptibility at zero frequency looks very similar to the 
spin counterpart. Originally it comes from the nesting properties
of the Fermi surface resembles by $\chi_0$. However, the 
additional enhancement due to the denominator is quite 
different origin in both cases. In the spin susceptibility it results due to
superexchange interaction having a maximum at ($\pi$, $\pi$). In the
charge susceptibility the contribution from $J_q$ has a different sign
and therefore should suppress the peak. On the other hand inclusion of the 
inter-site Coulomb screening repairs the situation and again leads to the 
strong commensurate peak around ($\pi$, $\pi$). As we will see later this 
peak is strongly dependent only on the value of inter-site Coulomb repulsion.
This is quite remarkable fact if one remembers that 
for example the calculations
of the charge and spin susceptibility in the frame of one-band Hubbard
Hamiltonian with on-site $U$ or ordinary $t-J$ model show that 
charge susceptibility is
\begin{figure}
\begin{center}
%\centerline{\epsfig{clip=,file=fig1.eps,width=10.7cm,angle=-90}}
\hspace{0cm}{{\psfig{figure=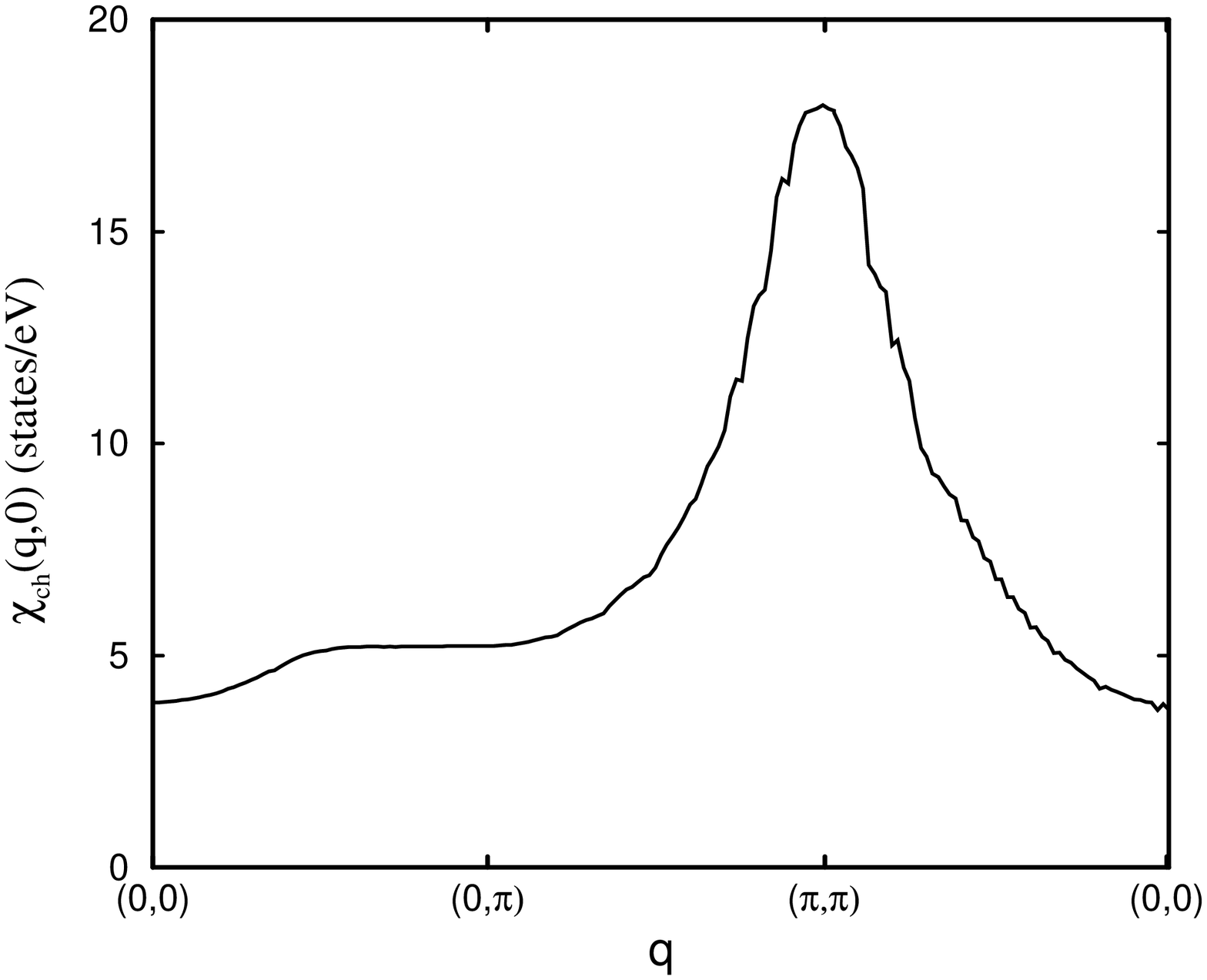,width=7cm,angle=-90}}}
\end{center}
\vspace*{-1.0cm}
{\small FIG. 1. Momentum dependence of the Re $\chi ({\bf q}, 0)$  
through the route of the Brillouin zone $(0,0)-(\pi, 0)-(\pi,\pi)-(0,0)$ at
T=100 K.}
\end{figure}
rather small at Q=($\pi$, $\pi$) in contrast to the spin 
susceptibility\cite{scalapino}. This often led to the exclusion of the charge
degrees of freedom from the pairing interaction. However, as it is seen here
the significant contribution to the charge susceptibility comes from the
inter-site screened Coulomb repulsion between doped holes. Since this
interaction plays the most important role in underdoped case one would expect
the significant contribution to the pairing interaction from the
charge susceptibility too.

This fact is also seen from Fig. 2 where 
we present the imaginary part of the charge susceptibility at 
Q=($\pi$,$\pi$) with and without screened Coulomb repulsion $G_1$. 
Without inter-site Coulomb repulsion one sees that charge susceptibility 
shows no features at low energies. This indicates that superexchange
mechanism itself cannot lead to any charge instabilities in the system.
The situations changes drastically if one switches on the screened Coulomb 
repulsion between doped holes. At $G_1$ = 70 meV charge susceptibility 
shows a strong peak at $\omega_{cf}$ = 15 meV. This is quite
comparable with typical spin fluctuation frequency, $\omega_{sf}$ 
that results from the approximate position of the
\begin{figure}
\begin{center}
\hspace{0cm}{{\psfig{figure=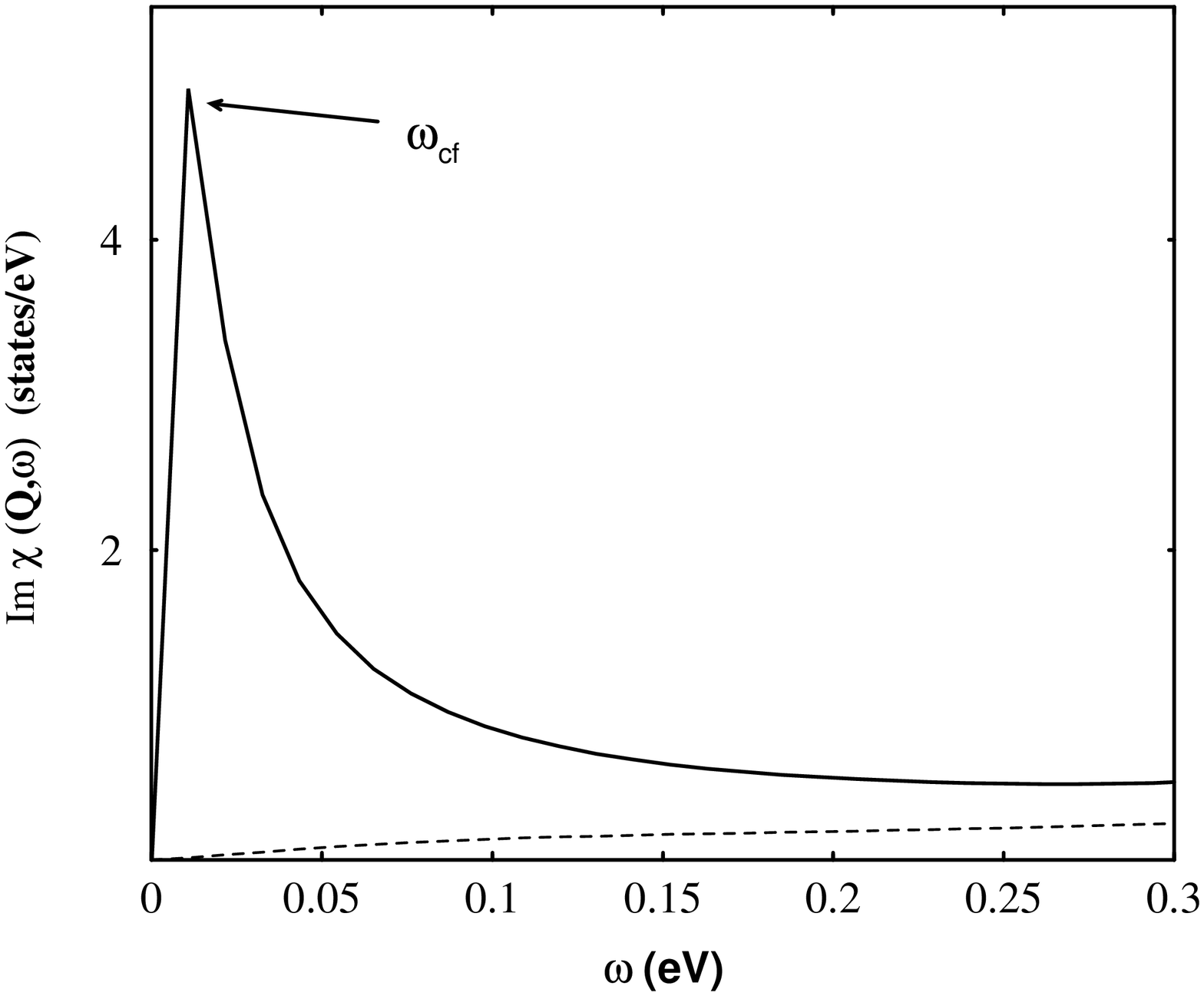,width=7cm,angle=-90}}}
\end{center}
\vspace*{0.2cm}
{\small FIG. 2. Frequency dependence of the Im $\chi ({\bf Q}, \omega)$  
at Q = ($\pi$, $\pi$) for G$_1$=0 (long-dashed curve) and G$_1$= 70 meV
(solid curve) at T = 100 K. The arrow indicates the approximate position 
of the plasmon or charge fluctuation frequency $\omega_{cf}$ $\approx 15 meV$.}
\end{figure}
peak in Im $\chi_s$, $\omega_{sf}$ and which is also in order of 15 meV
in layered cuprates\cite{mmp}. On the basis of Figs. 1 and 2 
we can conclude that
charge susceptibility may play an important role in the scattering mechanism
and in the Cooper pairing instability correspondingly. 
However, such a large
value of inter-site Coulomb interaction  one could expect  only at 
low doping level. Therefore, we think that the spin fluctuation scenario of 
superconductivity is much more robust than charge 
fluctuation one in high-T$_c$ cuprates. One would expect, however, a strong
charge density wave like instability in underdoped cuprates since inter-site
Coulomb repulsion is the most important at low doping concentration. 

In order to analyze these possible instabilities in the charge
subsystem we also investigate the charge susceptibility with respect
to a CDW formation. The CDW frequencies range can be obtained analyzing the 
denominator of dynamical charge susceptibility expression (\ref{cds}). 
Therefore, we solve the equation
\begin{equation}
1+\lgroup G_q -\frac{J_q}{4}\rgroup \chi_{0}({\bf q}, \omega) + 
\chi_{1}({\bf q}, \omega) - \lgroup 1-\frac{P_{pd}}{2}\rgroup 
z({\bf q}, \omega) = 0
\label{zqw}
\end{equation}
through the route of the Brillouin zone $(0,0)-(\pi, 0)-(\pi,\pi)-(0,0)$.
One has to notice, however, that the solution of Eq. (\ref{zqw}) can only be 
obtained at larger than 70 meV values of inter-site Coulomb 
repulsion. In particular,
at the optimal doping, $i.e.$ $p=0.16$, the stable solution in the whole
Brilloune zone was found only at G$_1$ = 150 meV. In Fig.3 we present
the resulting energy dispersion of CDW frequency, $\omega_q^{CDW}$ 
at temperature T=100 K.
As one can see there are three disconnected branches of the frequency curves.
It happens due to the fact that instability occurs not along the  
whole Brllouin
zone but rather at some critical wave vectors. In particular, 
it is clearly  seen that
the charge system is very unstable near the incommensurate wave vector
($2/\pi/3$, $2\pi/3$) where curve is crossing zero. This means that
the possible CDW must occur at this incommensurate wave vector 
${\bf Q}_{CDW} \approx (2\pi/3,2\pi/3)$. This quite a remarkable result  
confirms the difference between paramagnon charge fluctuations which are
commensurate and real CDW order parameter having incommensurate wave vector.
\begin{figure}
\begin{center}
\hspace{0cm}{{\psfig{figure=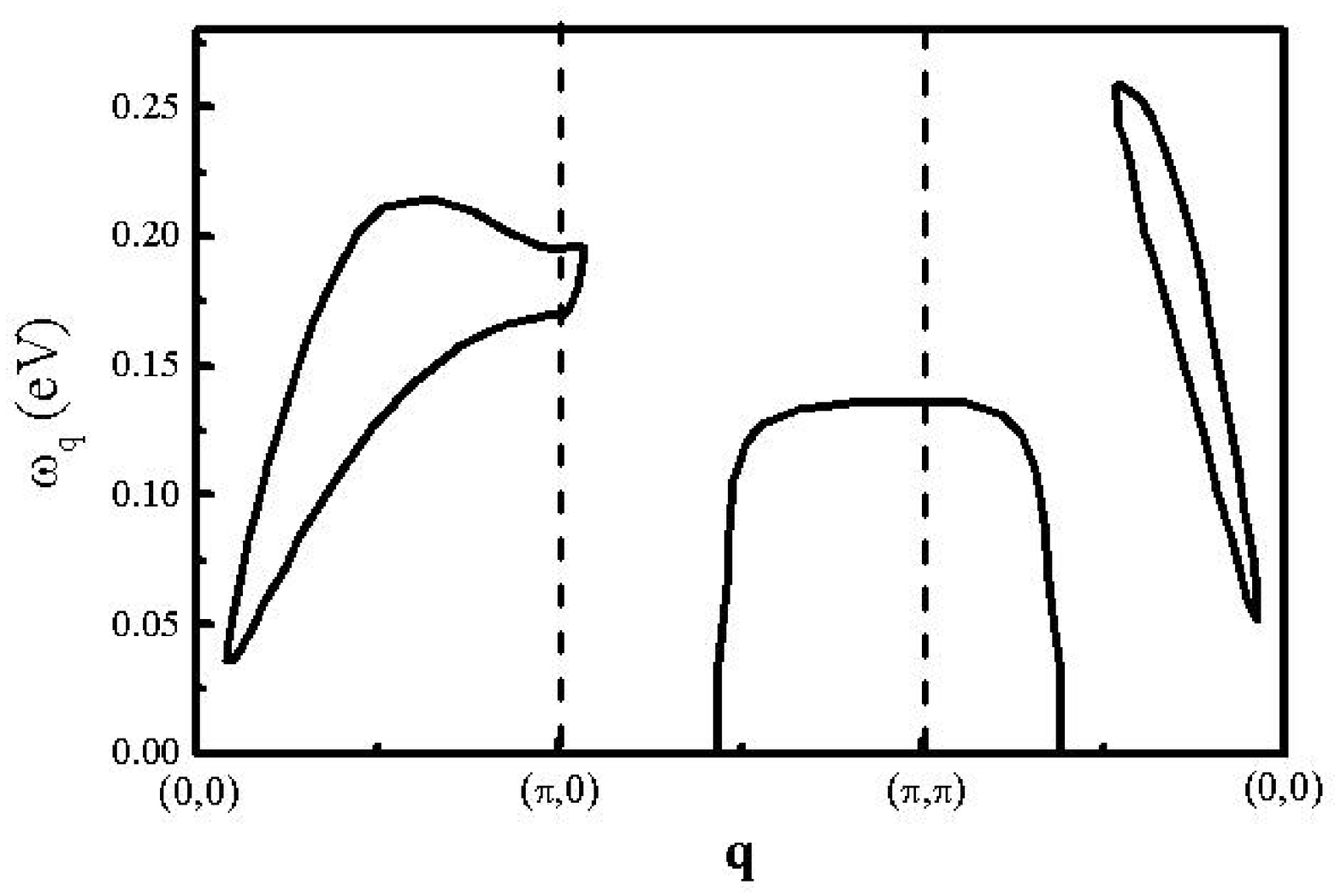,width=8cm,angle=0}}}
\end{center}
\vspace*{0.2cm}
{\small FIG. 3. Momentum dependence of the CDW frequency acoustic mode
at G$_1$= 150 meV.}
\end{figure}

\section{Conclusion} 

In a conclusion we have obtained in a frame of singlet-correlated band model
beyond RPA approximation the new analytical expression 
for the dynamical charge sucseptibility.
We have shown that it produce a strong peak
at {\bf Q} = ($\pi$, $\pi$) at large enough screened Coulomb repulsion
between doped holes. This may result in a significant contribution 
to the pairing interaction from the charge susceptibility at low doping level
where Coulomb repulsion plays an important role. 
The analysis of the instability of the system with respect to 
a CDW formation shows that the system is the most unstable at the
incommensurate wave vector {\bf Q}$_{CDW} \approx$ ($2\pi/3$, $2\pi/3$).

Its pleasure to thank V. Yushankhai, D. Manske, and G. Seibold
for stimulating discussions. This work is supported by the 
Russian Scientific Council on Superconductivity (Grant No. 98014)
and partially by the Swiss National Foundation (Grant No. 7SUPJ062258).
One of us (I.E.) would like to thank the Alexander von Humboldt
Foundation for the financial support.

\end{multicols}

\end{document}